\documentclass{PoS}
\usepackage{graphicx}
\usepackage{comment} 
\usepackage{rotating}
\usepackage{epsfig}
\usepackage{amssymb,amsmath,amsfonts}
\usepackage{bm}
\usepackage{paralist}
\usepackage{cite}
\usepackage{booktabs}
\usepackage{multirow}
\def\OMIT#1{{}}

\def\d{{\delta}}
\def\D{{\Delta}}

\def\k{{\kappa}}

\def\L{{\Lambda}}

\newcommand{\beq}{\begin{equation}}
\newcommand{\eeq}{\end{equation}}

\newcommand{\bea}{\begin{eqnarray}}
\newcommand{\eea}{\end{eqnarray}}

\newcommand{\bee}{\begin{enumerate}}
\newcommand{\eee}{\end{enumerate}}

\newcommand{\bef}{\begin{figure}}
\newcommand{\eef}{\end{figure}}

\newcommand{\bei}{\begin{itemize}}
\newcommand{\eei}{\end{itemize}}

\newcommand{\benn}{\begin{displaymath}}
\newcommand{\eenn}{\end{displaymath}}
\newcommand{\ket}[1]{| #1 \rangle}                     
\newcommand{\bra}[1]{\langle #1 \, |}                  

\title{Search for the H-Dibaryon in two flavor Lattice QCD}

\ShortTitle{Search for the H-Dibaryon in two flavor Lattice QCD}

\author{\speaker{Parikshit Junnarkar},
	Anthony Francis,
	Jeremy Green,
	Chuan Miao,
	Thomas Rae,
	Hartmut Wittig
	\\
	Helmholtz-Institut Mainz,
	Institut f\"ur Kernphysik,
	and PRISMA Cluster of Excellence,\\
	Johannes Gutenberg-Universit\"at Mainz, Germany\\
	Department of Physics and Astronomy, York University, Toronto Canada \\
	Bergische Universit\"at Wuppertal, Germany\\
	E-mail:
	\email{junnarka@kph.uni-mainz.de},
	\email{afranc@yorku.ca},
	\email{green@kph.uni-mainz.de},
	\email{chuan@kph.uni-mainz.de},
	\email{thrae@uni-wuppertal.de},
	\email{wittig@kph.uni-mainz.de}

}
	\abstract{
We present preliminary results from a lattice QCD calculation of the H-dibaryon using two flavors of $\mathcal{O}(a)$ improved Wilson fermions. 
We employ local six-quark interpolating operators at the  source with a combination of local six-quark and two-baryon operators at the sink with the appropriate quantum numbers of the H-dibaryon and its coupling to the two-baryon channels. 
We find that the two-baryon operators provide an improved overlap onto the ground state in comparison to the local six-quark operators. 
We also apply L\"uscher's finite volume formalism to obtain information on the nature of the infinite-volume interaction of two particles. 
Further, the momentum projection to three moving frames enables the isolation of the pole in the infinite-volume scattering amplitude. 
Preliminary results at pion masses of 450 MeV and 1 GeV clearly indicate the presence of states below the $\L \L$ threshold while a finite-volume analysis fails to conclusively show the existence of an infinite-volume bound state. 
	}

\FullConference{The 8th International Workshop on Chiral Dynamics, CD2015 ***\\
		29 June 2015 - 03 July 2015\\
		Pisa,Italy}

\begin{document}
\vspace{-0.9cm}
\section{Introduction}
The existence of a deeply bound six-quark state in the scalar SU(3) singlet channel ($J^P=0^+,\ S=-2$), the so called H-dibaryon, was first proposed by Jaffe \cite{Jaffe:1976yi} in 1977 using the MIT bag model. 
Experimentally such a deeply bound state is ruled out and the strongest constraint is provided by the KEK experiment E373~\cite{Takahashi:2001nm} known as the ``Nagara'' event limits the H-dibaryon to be loosely bound.
Recent lattice QCD calculations by NPLQCD \cite{Beane:2010hg} and HALQCD \cite{Inoue:2010hs} indicate that at heavy pion masses the H-dibaryon is bound. 
In our previous work~\cite{Green:2014dea,Francis:2013lva}, it was not clear that the H-dibaryon was bound even at $m_\pi = 1 $ GeV, motivating the expansion of the operator basis to include  multi-baryon operators. Here we present preliminary results from such an expanded operator basis.  
\section{Lattice set-up}
\vspace{-0.3cm}
The ensembles used in this work employ non-perturbatively $\mathcal{O}(a)$ improved Wilson fermions in $N_f = 2 $ QCD and were generated as a part of the CLS effort.
In order to compare results between local six-quark and two-baryon operators, we choose the same ensembles used in our previous work namely, Ensemble E1 ($m_\pi = 1 $ GeV) and Ensemble E5 ($m_\pi = 450$ MeV).
The relevant ensemble parameters were also set identical to the ones in our previous work and have been described in Ref.~\cite{Green:2014dea}.
On  E5, the number of measurements was doubled in order to have an unambiguous isolation of the ground state. 
To improve the overlap onto to the ground state, Wuppertal smearing~\cite{Gusken:1989qx} was applied on quark fields with spatially smeared APE links. 
In our previous calculation, we employed three smearings each with a different number of smearing steps as narrow ($N = 70$), medium ($N = 140$) and wide ($N = 280$) and found that the wide smearing to be ineffective due to significant smearing noise.
As a consequence in this work we choose to retain only the narrow and medium smeared quark fields. 

In the work presented, we employ local six-quark positive-parity-projected interpolating operators at the source and sink and two-baryon operators at the sink. 
The definition and construction of the local six-quark operators remain unchanged and the reader is referred to Ref.~\cite{Green:2014dea} for the relevant details.  

For the two-baryon operators, there are three possible combinations of octet baryons with quantum numbers ($S=-2,I=0$).
In the flavor basis, they are:
\bea
(\Lambda \Lambda) &=& \frac{1}{12}[sud][sud], \quad (N \Xi) = \frac{1}{18 \sqrt{2}} ( [uud][ssd] - [dud][ssu]), \\
(\Sigma \Sigma) &=& \frac{1}{36 \sqrt{3}}(2[uus][dds] - [dus][uds] - [dus][dus] - [uds][dus] - [uds][uds] + 2[dds][uus] ), \nonumber 
\eea 
where a three-quark operator with flavor content $[abc]$ is constructed as $[abc]= \epsilon^{ijk}(b^T_i C \gamma_5 P_+ c_j)a_{k}$. In practice, it is useful to transform the aforementioned states to the SU(3) basis where they belong to the singlet, octet and 27-plet irreps. 
In computing the two-point function, a momentum projection of the two-baryon states $(B_1B_2)$ at the sink with definite momenta $\vec{p_1}$ and $\vec{p_2}$ is schematically done as follows:
\beq
B_1B_2(\vec{p}_1,\vec{p}_2) = \sum_{\vec{x},\vec{y}} e^{i \vec{p_1} \cdot \vec{x}} e^{i \vec{p_2} \cdot \vec{y}} \ B^T_1(\vec{x}) \ C \gamma_5 P_+ \ B_2(\vec{y}), P^2 = ( \vec{p}_1 + \vec{p}_2 )^2 = \bigg( \frac{2 \pi}{L} \bigg)^2 \big\{0,1,2,3 \big \}. 
\eeq
In performing the momentum projection, the maximum individual component of the momenta $\{P,p_1,p_2\}$ is set to $\frac{2 \pi}{L}$ which allows for projection onto four frames with total frame momentum $P^2$ as indicated. 
In a given frame, we retain different two-baryon operators with a given total momentum $P^2$ and different $(p_1^2, p_2^2)$ and use them to explore the ground state as described in the next section.
\section{Results}
We compute a matrix of correlators with the interpolating operators described in the previous section. For a given set of source operators $\mathcal{O}_j$ and sink operators $\mathcal{O}_i$, we compute,
\beq
C_{ij}(t) = \bra{0} \mathcal{O}_i(t)  \mathcal{O}^{\dagger}_j(0) \ket{0}, \quad C_{ij}(t+\Delta t) v_j(t) = \lambda (t) C_{ij}(t) v_j(t)
\eeq
and solve a generalised eigenvalue problem (GEVP)~\cite{Blossier:2009kd}. 
The effective masses are computed from the eigenvalues as $m_{\mathrm{eff}}(t) = - \frac{\mathrm{log} \lambda (t)}{\Delta t}$.

We now describe the choices of the operators available to us for solving the GEVP. 
On  E1, due to the SU(3) symmetry, there is no mixing between singlet, octet and 27-plet irreps and combined with the two smearings, there are two operators available for each multiplet at the source. 
At the sink, we choose to include only the narrow smeared operators as they are less noisy. 
Hence, there is a choice of one local six-quark operator and different two-baryon operators with a given total $P^2$.\footnote{There are four different operators in $P^2=0$ frame correpsonding to $p_1 = -p_2$, $p_1^2=(2\pi/L)^2\{0,1,2,3\}$ while in frames $P^2=(2 \pi/L)^2\{1,2,3\}$ there are different number of operators. 
In moving frames, parity not being a good quantum number allows for additional operators. Those operators are not included in this analysis for simplicity.}
We also note that since there are no two-baryon operators at the source, the operator basis of the GEVP is asymmetric and consequently the correlator matrix is non-hermitian. 
We explore the ground state of this system by analysing various square correlator matrices. 
\bef[t!]
\centering
\includegraphics[width=1.0\linewidth]{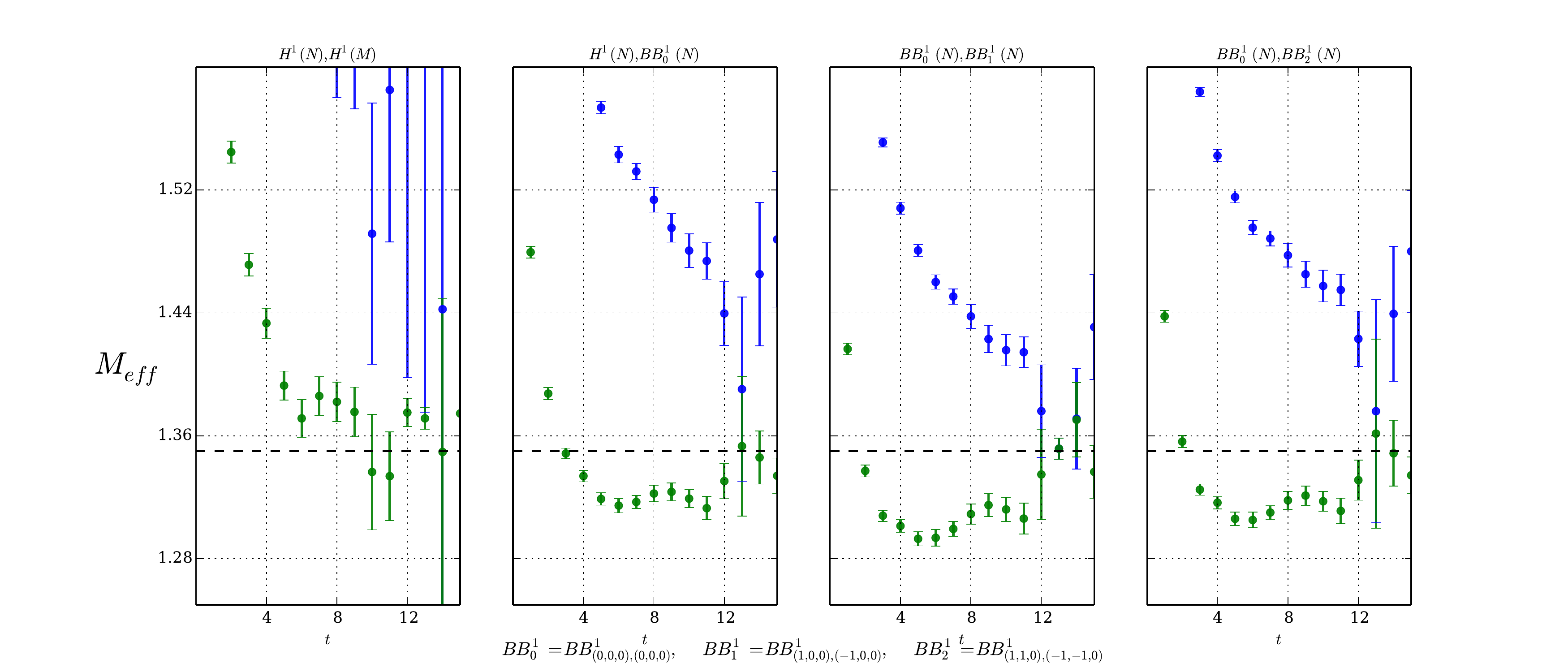}
\caption{\label{fig:E1-H1} Results of GEVP with various sink operators in the H-dibaryon channel on Ensemble E1. Dashed line corresponds to $2 m_\L$ level. Results from panel two are used in the finite-volume analysis. For a description of the panels, see text.}
\vspace{-0.25cm}
\eef
\bef[t!]
\centering
\includegraphics[width=1.0\linewidth]{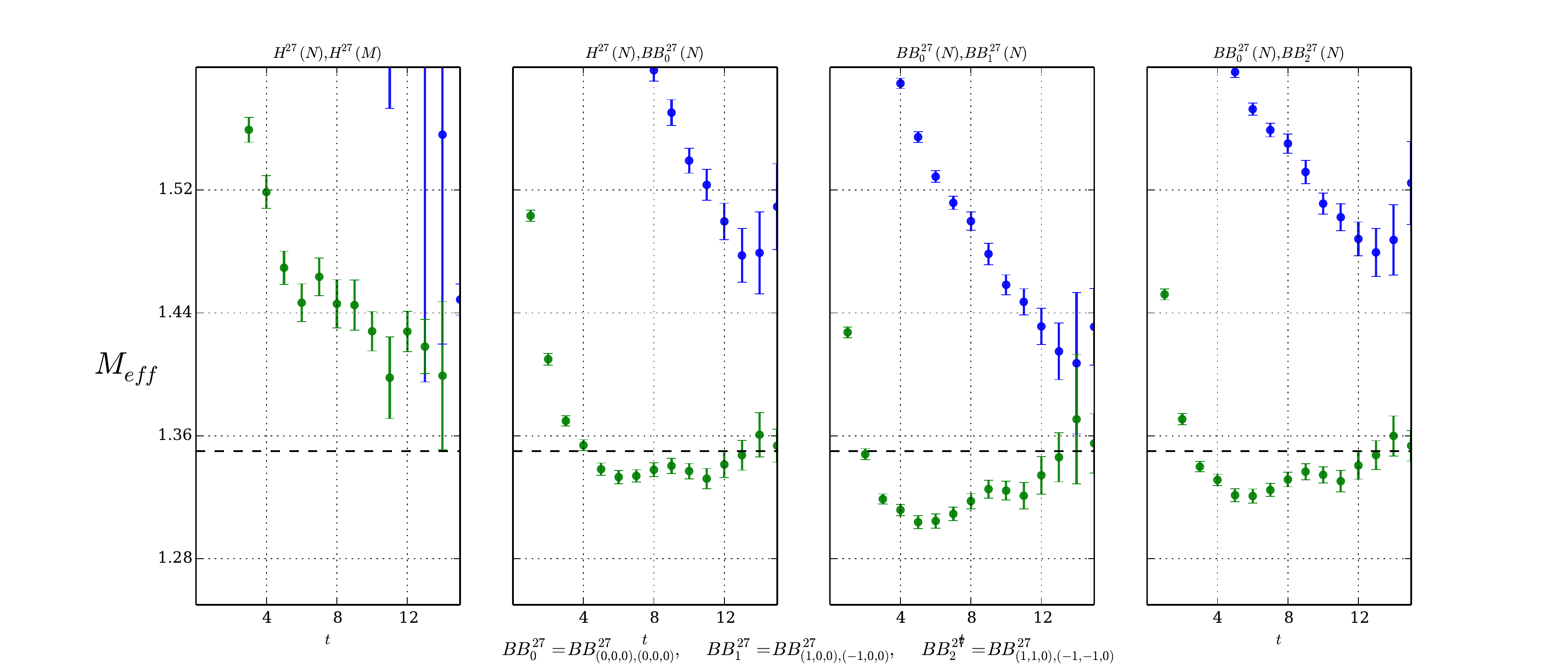}
\caption{\label{fig:E1-H27} Results of GEVP with various sink operators in the 27-plet channel on Ensemble E1. Results from panel two are used in the finite-volume analysis}
\vspace{-0.25cm}
\eef
\bef[t!]
\centering
\includegraphics[width=1.0\linewidth]{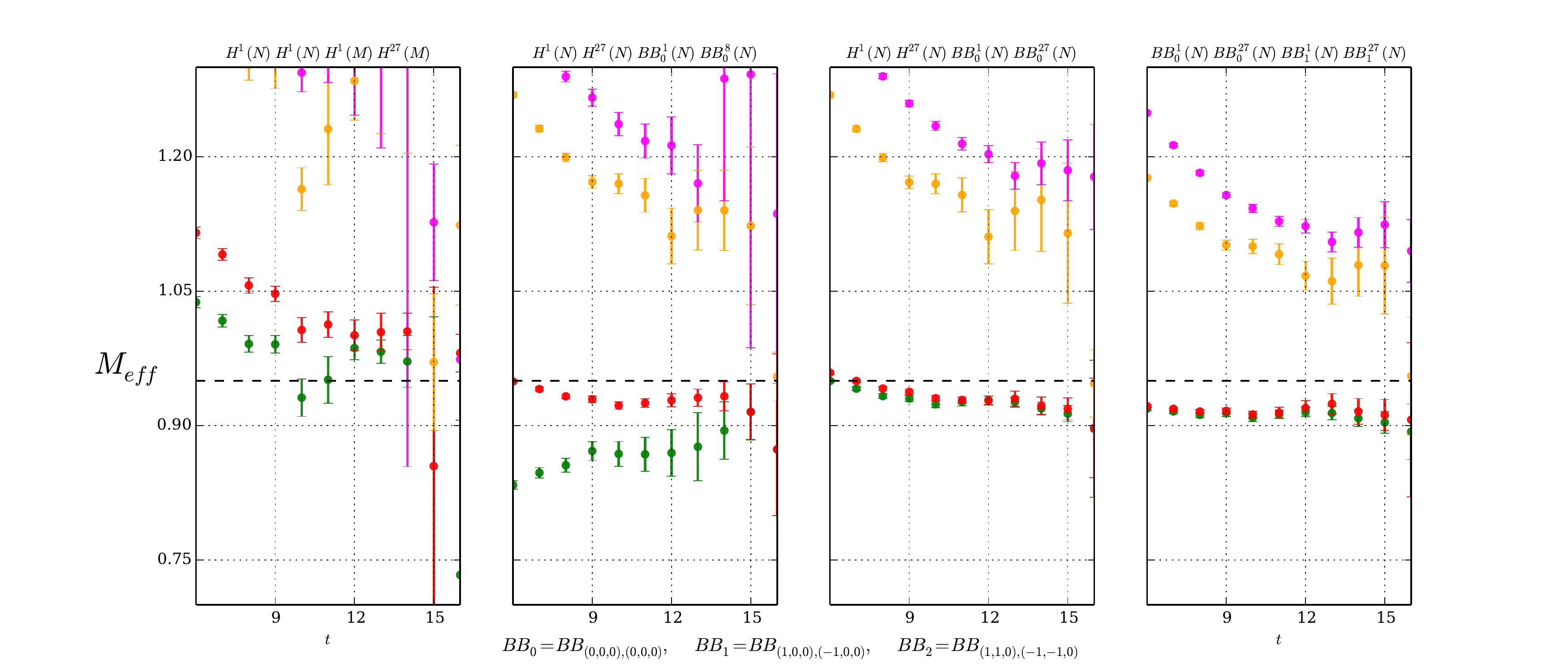}
\caption{\label{fig:E5} Results of GEVP with various sink operators Ensemble E5. Results from panel three are used in the finite-volume analysis}
\vspace{-0.25cm}
\eef
The results for the SU(3) singlet are presented in Fig.~\ref{fig:E1-H1}. 
The different panels show results from solving the GEVP with different sink operators from the choices described above. 
The first panel shows the results with using only local six-quark source and sink operators and it provides no clear evidence for the presence of a bound state. 
In the second panel, we used a combination of a local six-quark operator and a two-baryon operator projected to the $P^2=0$ frame and there is clear indication of the presence of a bound state. 
In panels three and four, we use various two-baryon operators with $P^2=0$ which also clearly indicate the presence of a bound state. 
A similar GEVP analysis is performed on operators projected to total momenta of $P^2=\big(\frac{2 \pi}{L} \big)^2 \{1,2,3\}$.    

The results for the 27-plet are presented in Fig.~\ref{fig:E1-H27}. 
The available choices of operators to construct various GEVP's, with the relevant quantum numbers of 27-plet, are the same as those of the SU(3) singlet. 
The presence of a bound state is unclear when only local six-quark operators are employed at the sink.
The inclusion of two-baryon operators either in combination with a six-quark operator or with different two-baryon operators with given total $P^2$ clearly indicates the presence of a  bound state.

On  E5, the breaking of SU(3) symmetry leads to mixing between the singlet, octet and 27-plet irreps.
This requires an enlarged operator basis for exploring the ground state with different GEVP's.
At the source, there are four operators available, namely, $H^{\mathbf{1}}$ and $H^{\mathbf{27}}$ with two different smearings, which sets the size of the square correlator matrix to $4\times4$.
At the sink, as before using only narrow smeared operators, we have two local six-quark operators ($H^{\mathbf{1}}$ \& $H^{\mathbf{27}}$) and different two-baryon operators with a given total $P^2$ in the singlet, octet and 27-plet irreps. 
The results with various sink operator choices are shown in Fig.~\ref{fig:E5}. 
In the first panel, as before, we employ only local six-quark operators and it is not clear whether a bound state exists.
In panel two, we employ a combination of local six-quark and two-baryon operators including an octet two-baryon operator which is only present at the sink. 
The results clearly indicate the presence of two bound states with a state approaching from below coming from the octet-baryon operator. 
Since there is no octet operator at the source, it is difficult to resolve this state and we defer the issue to future studies.
Such work is currently underway and this operator is not included in the further analysis.
In panels three and four, we use combination of six-quark and different two-baryon operators with a given total $P^2$ such that the SU(3) irreps at the source and sink are same.
In both cases, the results indicate the presence of two bound states below threshold however due to insufficent statistics we are unable to isolate the energy difference between the states.
We note that this is in contrast to the findings of Ref.~\cite{Beane:2010hg} where the possiblilty of two states below threshold was discounted.
\vspace{-0.2cm}
\section{Finite volume analysis}
\vspace{-0.2cm}
\bef[t!]
\hspace{-1.0cm}
\includegraphics[width=0.55\linewidth]{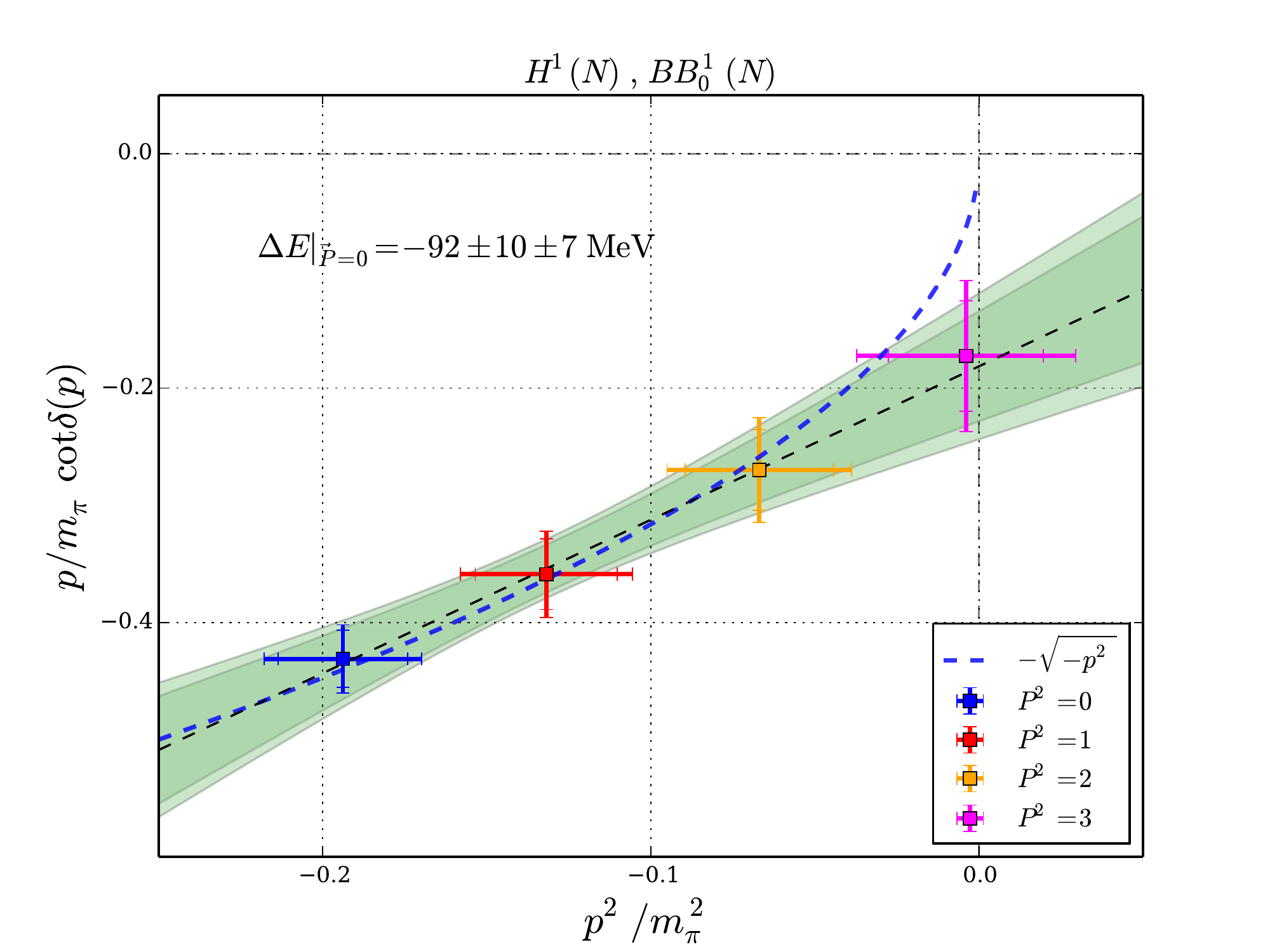}
\includegraphics[width=0.55\linewidth]{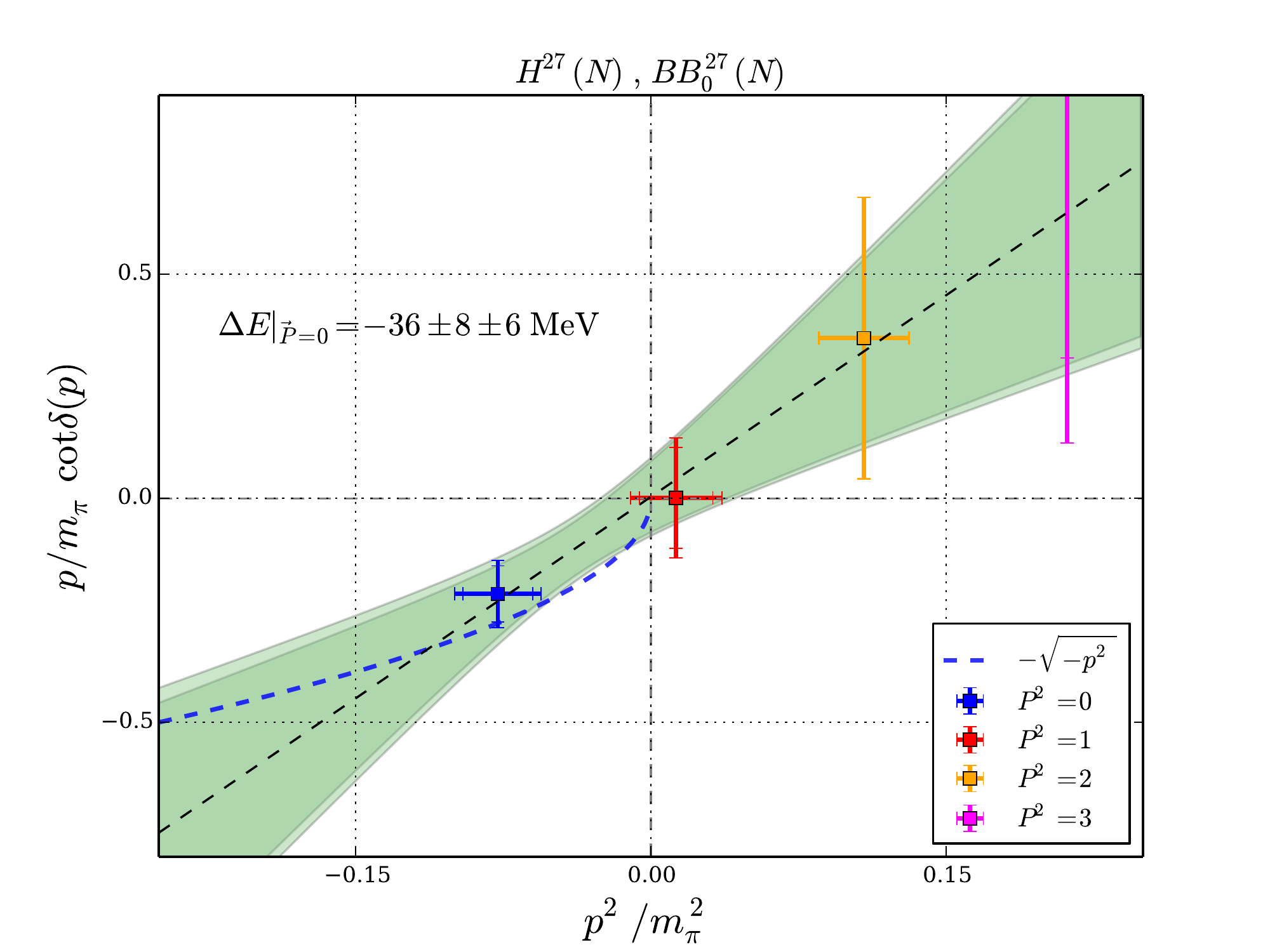}
\caption{\label{fig:FV-E1} Left: $p \ \text{cot} \d$ for the SU(3) singlet (H-dibaryon) channel on Ensemble E1. Right: $p \ \text{cot} \d$ for the 27-plet channel on Ensemble E1. Please refer to the text for a description of fits. The quoted binding energy is computed as $\D E = E_{\L \L} - 2 M_{\L}$ in the rest frame. The first quoted uncertainty is statistical while the second one is systematic. See footnote 2 for the description.}
\vspace{-0.25cm}
\eef
\bef[t!]
\centering
\hspace{-1.0cm}
\includegraphics[width=0.55\linewidth]{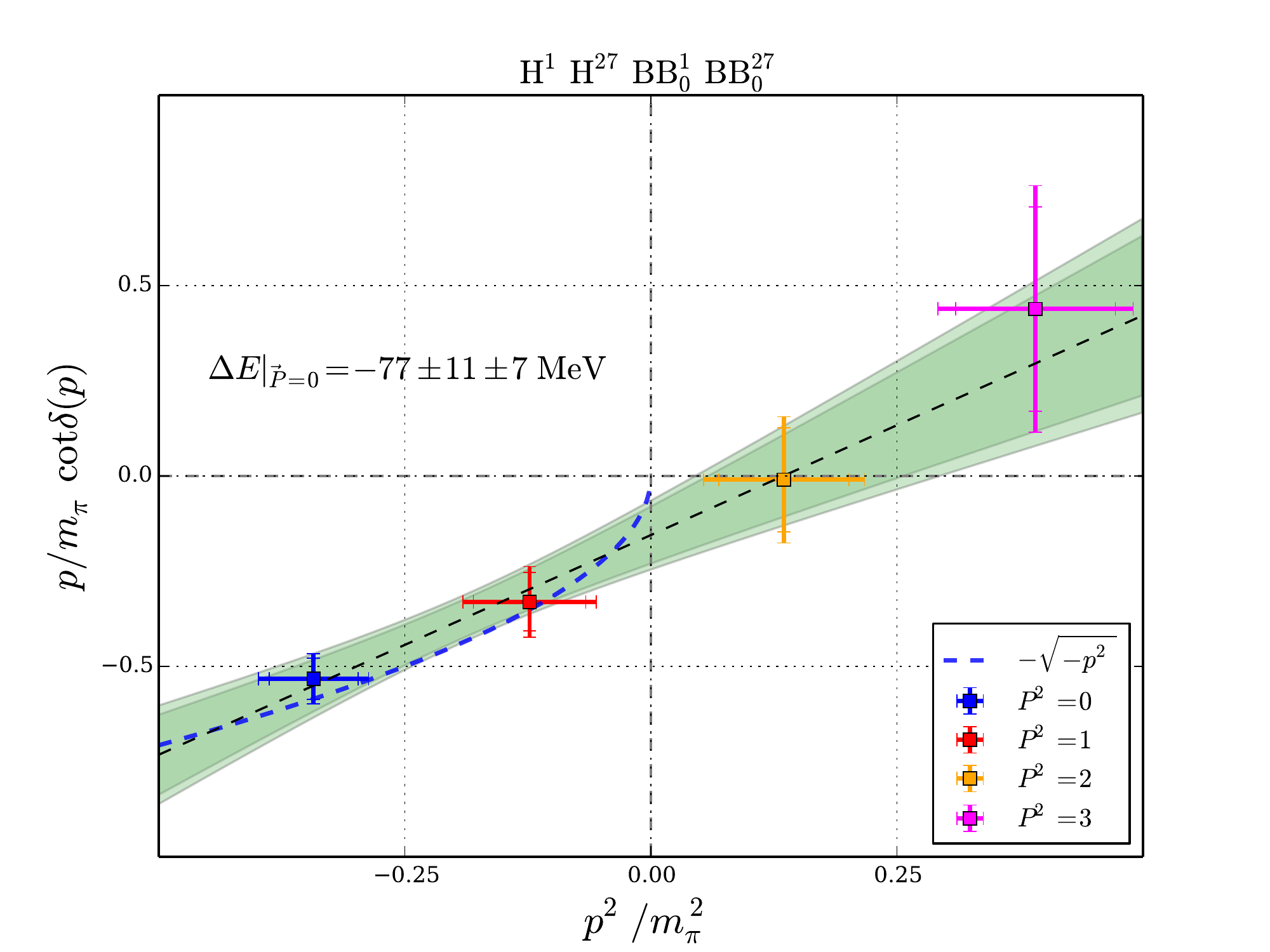}
\caption{\label{fig:FV-E5}Results for $p \ \text{cot} \d$ on Ensemble E5. The quoted binding energy is computed as $\D E = E_{\L \L} - 2 M_{\L}$ in the rest frame.}
\vspace{-0.25cm}
\eef
The results for the ground states, as described in the previous section, can be used to study the infinite-volume two-particle scattering from L\"uscher's finite-volume formalism \cite{Luscher:1990ux} and its extension to moving frames~\cite{Rummukainen:1995vs}. 
For a given ground-state energy $E$, obtained by fitting the effective mass in the plateau region,~\footnote{In addition to estimating the statistical uncertainty originating from finite number of Monte-Carlo gauge configurations, a systematic uncertainty due to fitting over finite fitting range, is estimated by performing correlated fits over several fitting ranges.}the associated binding/scattering momentum of the single particle is given as, $p^2 = \frac{1}{4}\big(E^2 - P^2 \big) - M^2$, where $M$ is the ground state mass of the single particle. 
L\"uscher's formalism~\cite{Luscher:1990ux} provides a mapping of this momentum to the infinite-volume scattering phase-shift and in the rest frame is given as,
\beq
p \ \text{cot} \delta(p) =\frac{2}{\sqrt{\pi}L} \mathcal{Z}_{0,0}(1;q^2), \quad \mathcal{Z}_{0,0}(1;q^2) = \frac{1}{\sqrt{4 \pi}} \lim_{\L \to \infty}\bigg\{ \sum^{\L}_{|\vec{n}|} \frac{1}{q^2 - n^2} - 4 \pi \L\bigg\},
\eeq
where $q = \frac{pL}{2 \pi}$. In the moving frames, the above relation for the zeta function $\mathcal{Z}_{0,0}(1;q^2)$ has to be  modified and one possible derivation, used here, was given in Ref.~\cite{Gockeler:2012yj}. 
The scattering phase-shift thus computed uniquely specifies the associated two-particle scattering amplitude which has a general form given by $ \mathcal{A} \propto 1/(p \ \text{cot} \delta(p) - ip)$.
The pole in this scattering amplitude indicates the presence of a bound state with the binding momentum specified by the location of the pole. 
In order to locate the pole, we fit our data set to the effective range expansion (ERE)\footnote{The momenta employed in this work are well below the t-channel cut justifying the validity of ERE.} given as $p \ \text{cot} \d(p) = - 1/a + (r p^2)/2 + \ldots $ including the first two terms, where $a$ denotes the scattering length and $r$ denotes the range parameter.
The fitted results are then compared with the binding momentum to locate the pole.   

In Fig.~\ref{fig:FV-E1}, we present results of the finite-volume analysis on  E1 where the computed values of $p \ \text{cot} \delta(p)/m_\pi$ are plotted against the binding momentum $p^2/m^2_\pi$. 
The momentum dependence of the results arises due to the projection of the correlators onto different frames as indicated in the legend of the figure.
The fit to the ERE is represented by the black dashed line and the inner green band indicates statistical uncertainty while the outer band indicates a total of statistical and systematic uncertainty on the fit.
The blue dashed line corresponds to the binding momentum $ \k = -\sqrt{- p^2}$ whose intersection with the ERE fit locates the pole in the scattering amplitude.
In the H-dibaryon channel (SU(3) singlet) as shown in left panel of Fig.~\ref{fig:FV-E1}, the ground state is deeply bound in the rest frame and the ERE fit is almost tangential to the binding energy curve $\k$. 
The uncertainty on the fit overlaps significantly over the binding energy curve, making the location and the existence of the pole ambiguous.
The results in the 27-plet channel also exhibit similar trends where in the rest frame we have a shallow bound state and the size of the uncertainty on the fit makes it difficult to locate the pole precisely.  
On  E5, we find the ground state to have a more shallow binding energy compared to  E1 and the finite volume analysis shown in Fig.~\ref{fig:FV-E5} fails to locate the pole corresponding to the infinite-volume bound state. \\

{\noindent \bf Conclusions:}
\noindent
In the work presented here, we find that the use of two-baryon operators clearly show the presence of states below the $\L \L$ threshold, suggesting a bound H-dibaryon,  at both pion masses of $m_\pi = 1 \ \text{GeV} $ and $m_\pi = 450 $ MeV.
In the infinite-volume, we find that our data is well described by the effective range expansion.
However it does not allow for a clear isolation of the pole in the scattering amplitude at both pion masses. 
Our current set-up does not allow for a symmetric operator basis and we are unable to clarify contributions to the ground state coming from operators whose quantum numbers are only present at the sink.
We expect the inclusion of two-baryon operators at the source will provide more insight in clarifying such contributions, and lead to a more reliable extraction of the spectrum. \\ 

{\noindent \bf Acknowledgements:}
\noindent
Our calculations were performed on the BG/Q ``JUQUEEN" computer at NIC, J\"ulich through the computer time allocated to the account HMZ21. 
We gratefully acknowledge this support. 
The HPC cluster ``Clover" at the Helmholtz-Institut Mainz was used for testing purposes. 
We are grateful to Dalibor Djukanovic for the technical support. 
We also thank Maxwell T. Hansen for many useful conversations.
We are grateful to our colleagues within the CLS initiative for sharing ensembles.

\bibliographystyle{JHEP-2-notitle}
\bibliography{dibaryon-cd15}

\end{document}